\documentstyle[12pt]{article}

\newlength{\dinwidth}
\newlength{\dinmargin}
\newcommand{\resection}[1]{\setcounter{equation}{0}\section{#1}}

\begin{document}
\def\lq{\left [}
\def\rq{\right ]}
\def\qq{Q^2}
\def\dmu{\partial_{\mu}}
\def\dmus{\partial^{\mu}}
\def\AA{{\cal A}}
\def\BB{{\cal B}}
\def\Tr{{\rm Tr}}
\def\gp{g'}
\def\gs{g''}
\def\ggs{\frac{g}{\gs}}
\def\mpp{m_{P^+}}
\def\mpm{m_{P^-}}
\def\mpt{m_{P^3}}
\def\mpz{m_{P^0}}
\newcommand{\be}{\begin{equation}}
\newcommand{\ee}{\end{equation}}
\newcommand{\bea}{\begin{eqnarray}}
\newcommand{\eea}{\end{eqnarray}}
\newcommand{\nn}{\nonumber}
\newcommand{\dd}{\displaystyle}
\vspace*{3cm}
\begin{center}
  \begin{Large}
  \begin{bf}
BREAKING ELECTROWEAK SYMMETRY STRONGLY$^*~ ^{\dagger}$\\
  \end{bf}
  \end{Large}
  \vspace{8mm}
  \begin{large}
{\it N. Di Bartolomeo and R. Gatto}\\
  \end{large}
{\it D\'epartement de Physique Th\'eorique, Univ. de Gen\`eve}\\
  \vspace{5mm}
\end{center}
\begin{quotation}
%\vspace*{1cm}
\begin{center}
  \begin{bf}
  ABSTRACT
  \end{bf}
\end{center}
  \vspace{5mm}
\noindent
The problem of electroweak symmetry breaking is reviewed with discussion of
future relevant experimentation at LHC and $e^+e^-$ linear colliders. The
possibility of strong electroweak symmetry breaking is examined in more detail,
using the BESS (Breaking Electroweak Symmetry Strongly) model as a basis for
the discussion. The formal constructions are briefly presented and the possible
expectations at future colliders are summarized.
\end{quotation}
\vspace{1cm}
\begin{center}
UGVA-DPT 1994/04-846\\\
hep-ph/9404264\\\
April 1994
\end{center}
\vspace{1cm}
\noindent
$^*$To appear in the Memorial
Volume for Professor Robert Marshak, edited by E.C.G. Sudarshan, World
Scientific Publishing Company.\\
\noindent
$^{\dagger}$ Partially supported by the Swiss
National Foundation.
\newpage
\resection{Introduction}

It is a honour to dedicate this contribution to the memory of Robert E.
Marshak, a man who gave so much to our profession. Besides important scientific
work, such as his contribution to $V-A$ theory together with George Sudarshan,
Robert Marshak will be recalled for having during all his life
effectively operated for the progress of high energy physics, expecially
through promotion of international collaboration. The Rochester Conferences
were a substantial but not isolated aspect of the sense of service to the
community that Robert Marshak professed, as he was always dedicated to social
progress, to peace, and to cooperation.

The problem of electroweak symmetry breaking, studied under certain aspects and
from some points of view, will be the main subject of this contribution.
The standard model describes the elementary world in terms of quarks
and leptons and their gauge interactions. The gauge structure consists of
color, responsible for strong forces, with group structure $SU(3)_c$,
 and electroweak charges
from $SU(2)_L \otimes U(1)$ \cite{GWS}. Quarks and leptons
form three generations, $(u,d,e,\nu_e)$, $(c,s, \mu,\nu_{\mu})$,
$(t,b,\tau,\nu_{\tau})$.  The photon,
the $W$ and $Z$, and the gluon are the gauge bosons.

Within such a beautiful synthesis there is one aspect which is usually
regarded as perhaps uncomplete and unsatisfactory. It is the mechanism for
symmetry breaking, which is responsible for the masses of $W$ and $Z$.

The gauge couplings are uniquely fixed from the gauge principle. On the other
hand the gauge principle alone does not directly provide for an understanding
of the symmetry breaking and of the masses. At this point the conventional
approach is to introduce an elementary scalar field,
and an ad-hoc scale, expressed by its
expectation value on the vacuum. The gauge theory, to be  stressed again,
 does not
say something unique on the scalar sector; in a way, one should look at such a
sector as a kind of grafting operated on the gauge theory, though not
in contrast to the gauge principle itself.

The simplest but somewhat artificial way to operate the grafting is to
introduce one or more scalar fields. The predictive power of the gauge theory
is
greatly reduced; and one would expect in the simplest and standard formulation
a
massive neutral scalar particle, the so-called ``physical Higgs", at some yet
unknown mass. For a slightly more complicated structure, one would expect
two additional neutral and one charged (of both charges) scalars. So, even
within this simplest realization of the symmetry breaking mechanism,
in terms of elementary scalars, one at least expects a new, yet unseen,
particle, whose discovery would of course  be crucial.

The more satisfactory realizations, from a theory point of view, are however
more complicated. As such, they lead to the expectation that higher energy
accelerators, than the ones we can presently use, might reveal a new realm of
particles and interactions. A general, common, theoretical idea, is that our
parametrization in terms of scalar couplings (for instance scalar mass term,
scalar self-coupling, all the list of Yukawa couplings ) is fundamentally the
effective low energy manifestation
of a more complicated dynamics, with additional particles and
interactions. The mass scale related to the Higgs vacuum value provides then
for
 a valuable information for the scale where these new particles may be found.

The new dynamics may, for instances, have the form of a new strong interaction
\cite{techni}. A simplest idea is that of reusing our technical
knowledge from QCD, supplementing it with additional inputs (for instance,
extended technicolor \cite{extechni}).

The idea of a composite model would be more radical.In these models quarks and
leptons are composites.

Another intensively discussed
possibility is supersymmetry \cite{susy}, according to a number of
different formulations. New particles, such as superpartners, are then
introduced, and the Higgs picture emerges in a stabilized form
(non-renormalization result).

This stabilization is thought to be useful to avoid the
difficult theoretical problem of naturalness, which afflicts the simple
breaking scheme based on elementary scalar fields alone. In this sense the
discovery of a Higgs of light mass would still leave the simple-minded Higgs
picture in face of its naturalness problem. It would rather suggests
supersymmetry as the next theoretical alternative, suggesting additional Higgs,
superpartners, etc.

Confirmation of the supersymmetry concept
would essentially come from the discovery of
the massive superpartners. Also, all supersymmetry schemes have at least two
doublets
of Higgs. Thus charged Higgs and more neutral ones would be expected in these
schemes.

In the standard model with a single scalar
doublet one has a theoretical lower mass
bound for the Higgs mass. The limit arises from the requirement that the Higgs
potential (as calculated at one loop ) be bound from below. In other words it
is a stability limit for the theory itself. For a top mass of $150~GeV$ such a
stability requires a Higgs mass larger than $92 \;GeV$. For a top of $200\;
GeV$ one would have instead a limit of $175 \; GeV$.

Other limits on the Higgs mass, again in the simplest standard model,  have to
do with perturbative unitarity. One requires that the amplitudes calculated in
perturbation theory, in practice at one-loop order or for leading logarithms,
satisfy unitarity. This is interpreted in terms of bounds for the Higgs mass.

Within the standard model, it is known that the interaction in the
scalar-longitudinal sector becomes stronger when the Higgs mass parameter,
$m_H$, of the scalar potential becomes very large \cite{VeltmanLee}.
Specifically, for large $m_H$, partial wave amplitudes for scattering among
longitudinal $W$, $Z$ and Higgs violate unitarity, when calculated at their
lowest order.

For a very massive Higgs one obtains that at energy higher than
$1.5 \; TeV$ unitarity (to be called strictly, perturbative unitarity)
 would be violated. Then either the Higgs is not very
massive, or the true amplitude is not the one calculated in perturbation
theory.
This relatively low value of $1.5 \; TeV$ has motivated much interest in higher
energy colliders (SSC, LHC).

A related theoretical speculation is to find out beyond which value of the
Higgs
mass the high enegy limit of $W W$ scattering violates perturbative unitarity.
In the standard model with one scalar doublet,
this happens for Higgs masses larger than $1.2 \; TeV$. On
the other hand at those masses the Higgs has a width already as large as its
mass. This can already be taken as an indication of failure of perturbation
theory.

In the approximation known as large N limit one
can construct an effective lagrangian which explicitly exhibits multiple
vertices between longitudinally-polarized $W's$ and $Z's$, together with
modified propagators, with unitarity visibly restaured \cite{3}.

The effective lagrangian constructed in such large N-model exhibits the scalar
Higgs resonance, whose properties coincide with those of the usual physical
Higgs for small $m_H$.This resonance
becomes broader for increasing $m_H$, and its mass
saturates at an upper bound of about 0.8 TeV \cite{4}.

A comprehensive
discussion of the phenomenology to be expected when the limit of failure of
perturbative unitarity is approached has been given by Chanowitz and
Gaillard \cite{5}, \cite{6}. A result, which relates the
amplitudes among longitudinally polarized gauge bosons to the corresponding
Goldstone amplitudes \cite{7}\cite{5}, is useful for these developments.

It is not yet known whether a large $m_H$ indeed generates a strong
interacting sector. The problem is theoretically difficult to solve
also for reasons connected with the possible triviality of $\varphi^4$.
The considerations which lead to the BESS (Breaking Electroweak Symmetry
Strongly) model \cite{bess} did in fact assume the existence of a strong
interacting longitudinal-scalar sector,
but they were not necessarily bound to the hypothetical
mechanism of large $m_H$. The model was rather constructed as a way of
parametrizing the most relevant phenomenological effects of a possible
strong interacting sector.

Complex poles, of various spin-parities, might be present in a complete
treatment of a strong interacting longitudinal-scalar sector, of whatever
origin.
Particularly interesting would be vector or axial poles. In fact, because
of their quantum numbers such poles could mix with the $W$ and $Z$
and thus originate visible deviations in the accessible phenomenology.

In ref. \cite{bess} the discussion of $J=1$ poles was made using an approach
which
goes back to the work by Callan, Coleman, Wess and Zumino \cite{9} on
non-linear
realizations of symmetry. The approach used the notion of hidden local
symmetry of non-linear $\sigma$-models, recently applied to describe vector
mesons in strong interactions \cite{10}.

The non-linear $\sigma$-model appears
in this context when one takes the formal limit of infinite $m_H$. By the
classical limit the isoscalar degree of freedom gets thus frozen. Beyond
this limit quantum fluctuations come in and the Higgs mass plays the role
of a cut off within the non-renormalizable non-linear $\sigma$-model \cite{11}.

Explicit gauge bosons
correspond to hidden local symmetries and
classically they appear as auxiliary fields. The physical hypothesis
here is that higher order effects provide for their kinetic terms,
as it happens in known two-dimensional examples \cite{12}. This is a
 hypothesis, and it simply adds to the many uncertainties in the
underlying dynamics (role of possible $\varphi^4$ triviality
\cite{13}\cite{14},
irrelevance of the $\sigma$-model limit at higher loop orders \cite{15},
various conjectures on fixed point mechanisms to prevent triviality
in Higgs sector \cite{16}, possible independence from the strength of the
quartic
coupling \cite{17}).

In view of the essentially unknown dynamics one may be
lead to consider the model developed in ref. \cite{bess} as an alternative
to the standard model in its realization of symmetry breaking.

One can
see, in fact, quite easily, that in general it is impossible to linearly
realize a spontaneous symmetry breaking mechanism in such a way that all scalar
degrees of freedom be eaten up via the Higgs mechanism.

In fact if the scalars
transform linearly under the gauge group (supposed connected, compact, and
semisimple) and if they all have to be absorbed, the invariant potential
could only be an overall constant,  being constant over a connected
compact subspace of same dimensionality as the representation.

When however
the symmetry breaking is realized through the mechanism of the
non-linear condition,
this conclusion can be avoided.
Note however that only geometric arguments
are used here and the possible plague of non-renormalizability is not taken
into
account.

In the standard model the scalar degrees of freedom would be three,
corresponding to the coordinates of the quotient space
$SU(2)_L \otimes SU(2)_R/SU(2)_{diagonal}$, the right number to give masses
to $W$ and $Z$. In the model of ref. \cite{bess} the hidden $SU(2)_V$ is
supposed
to have related gauge bosons $(V^{\pm}, V^0)$ and the quotient coordinates,
which are now six, are again all absorbed to give masses to $W,Z$ and $V$.

\resection{ Experimental prospects at future colliders}

Assuming the validity of
the one-doublet standard model
and leaving aside its probably unsatisfactory theoretical background, the
phenomenological discussion,
both for the past SSC project and for LHC \cite{Aachen}, has
focused on the possibility of detecting a heavy Higgs.
These studies provide at least a well defined
ground for exercises, particularly to formulate detector requirements.
The subject has been widely discussed in conferences and workshops. We shall
here present a rough overall picture, certainly not the final one.

Such a
heavy Higgs ( $m_H >0.5 \;TeV$ and up to $1 \; TeV$) would essentially decay
into $W W$ and twice less frequently into $Z Z$, with a total width
approximately given by $\Gamma_H (TeV)=1/2 (m_H/1 TeV)^3$, rather independent
of $m_t$. It would be produced mostly by gluon fusion and $W W$ or $Z Z$
fusion, with gluon fusion dominant at the lower masses, and (expecially for
small $m_t$ ) $W W$ and $Z Z$ fusion becoming dominant in the higher mass
range.

In the channel $ZZ \to 4 $ leptons, peaks would be present in the invariant
$m_{ZZ}$ mass. These peaks would be attributed to a heavy Higgs disintegrating
into  $ZZ$, with expected
branching ratio of the order of $4 \times 10^{-3}$. One
will have to cut on the low $Z$ transvers momenta and on the high $Z$
rapidities , and introduce some jet separation cut for the two jet process $qq
\to qq ZZ$.

A factor $\approx 6$ in the branching ratio could be gained by
observing the channel  $ZZ \to l^+ l^- \nu {\bar \nu}$, but in that case a
missing momentum cut excluding momenta lower than, say, $100 \; GeV$, will have
to be introduced. However, the less neat situation introduces a fake
background, due to $Z$ produced in association with jets simulating a missing
momentum.

To avoid the fake background one must demand that the hadron
calorimeter should cover a full range of hadronic rapidities up to four or five
units. For such channel, $ZZ \to l^+ l^- \nu {\bar \nu}$, one should look for
peaks in transverse $Z$ mass or transverse $Z$ momentum for the reconstructed
$Z$.

On such exercise-type heavy Higgs, the lower mass limit for the Higgs could
probably be pushed up to $0.6 \; TeV$ at LHC (we shall here refer to the
original project)
 with $10^4 \; pb^{-1}$ or
$0.8 \; TeV$ with $10^5 \; pb^{-1}$ within the $4\;l$ mode. Provided all
detector conditions be met, the lower luminosity using the $ll \nu \nu$ mode
might allow to go higher than $0.6 \; TeV$. All this is thinkable only provided
adeguate detectors for the high luminosity can be built.

The heavy Higgs case study we have just discussed should be regarded as the
typical study in order to define on a simple example the potentialities of
colliders. As we had discussed previously, the theoretical frame, for a
standard
model with heavy Higgs and nothing else, contains unconvincing features. One
may
than ask whether other similar exercises should not be pursued allowing for
different scenarios.

The simplest candidate for an alternative exercise is the
BESS-model. The model, in its simplest form,
predicts a new triplet of (strong-interacting) vector bosons $V^{\pm}$ and
$V^0$ of almost degenerate mass in the $TeV$ range. They would show up as
resonances in $pp \to W^{\pm}Z+X$, where $W^{\pm}$ could be seen in $W \to l
\nu$, and $Z$ in $Z \to ll$. The  $pp \to ZZ+X$  is non-resonant in BESS, and
$pp \to W^{+}W^- + X$  is dominated by $pp \to t {\bar t} +X$ with both
$t \to b W$.

The resonant signal would come from $q {\bar q} \to V \to
 W^{\pm}_L Z_L$ ( the suffix $L$ referring to longitudinal
polarization) and from
$W_L Z_L \to V \to W_L Z_L$, over a number of non-resonant contributions such
as continuum  $WZ$ production, $\gamma W \to WZ$, etc. Analysis of the
experimental possibilities, leads to a discovery
limit for $V$ of $\approx 2\;TeV$ in one year LHC (always the original project)
running at $10^{34} \; cm^{-2} \; s^{-2}$.

It is important in the whole context to also discuss the possibilities for
intermediate Higgs search, that is $m_H$ beyond the LEP2 limit up to the $2Z$
threshold. Here gluon fusion is the dominant $pp$ production process, and decay
is mostly into $b {\bar b}$, with small branching ratios into other modes. A
most effective possibility seems to be $pp \to ZZ^* X$
( $Z^*$ stands for virtual $Z$) with $ZZ^* \to 4l$ which
would work in the range $130-160 \;GeV$.

Other experiments would be $pp \to W H
X$ with $W \to l \nu$ and $H \to \gamma\gamma$, to possibly cover a lower mass
range, and, also $pp \to \gamma\gamma X$, based on the small $H \to \gamma
\gamma$ branching ratio, requiring very hard detector conditions.

In the
intermediate range, $e^+ e^-$ machines at adeguate energies and luminosities
($\ge 10^{32} \; cm^{-2} \;s^{-1}$) would have excellent prospects. The
intermediate $H$ would be looked at in its dominant mode $H \to b {\bar b}$.

For a high energy $e^+ e^-$, such as CLIC, one would look at the heavy Higgs in
$H \to WW$, $H \to ZZ$, and one expects for $30 \; fb^{-1}$ to reach limits
$\approx 0.5 \;TeV$.

A few words on supersymmetry.
As we have said in the previous section, supersymmetry remains
a valid theoretical
alternative. Supersymmetry can be formulated according to a variety of models.
Within each model many parameters are not fixed; calculations can be made for a
given set of parameters. This possibility of changing  the model and its
parameters does not unfortunately allow for universal predictions from
supersymmetry.

The so-called minimal supersymmetric standard model is in general used for a
first insight into supersymmetry, but it does not seem to possess fundamental
theoretical reasons to be preferred to other models.

Changing the model may
lead to brutal changes in the expected experimental signatures, as,
for instance,
if one goes to models where R-invariance does not hold  (R-invariance is
assumed to hold in the minimal supersymmetric standard model). One will have,
therefore, to qualify the meaning to be attributed to statements about
``reasonable" lower limits for the masses of supersymmetric partners that may
be reached at different future accelerators.

Supersymmetry predictions are at present object of intense studies.
At LHC in the main $pp$ option the search for squarks and gluinos appears as
the most promising among the superpartners searches, with expected
``reasonable" mass limits of the rough order of magnitude of $1\; TeV$ for both
$\tilde q$ and $\tilde g$ for a total $10^4 \; pb^{-1}$, and possibly $1.5 \;
TeV$ for $10^5 \; pb^{-1}$.

SSC would have allowed for still higher limits.
Tevatron, for comparison, could push the limit for $\tilde q$ and $\tilde g$ to
$0.2 \;TeV$. LHC in the $ep$ option would not compete with LHC-$pp$ on such a
limit. A $2 \;TeV$ $e^+ e^-$ with $500 \; fb^{-1}$ would not go beyond $0.8-0.9
\; TeV$ for this limit, but it would on the other hand allow to push up the
slepton mass lower limit, also to a similar value of $0.8-0.9 \; TeV$.

In this section we have given a rapid regard to some of the
possibilities that future accelerators can offer to look for standard Higgs,
even
up to rather heavy mass values, and we have given a quick assessment of
their possibilities with regard to supersymmetry, which is always considered to
be an interesting valid alternative to the standard model with some definite
theoretical advantages. In the present contribution
we shall essentially concentrate
on BESS (breaking electroweak symmetry strongly). We shall first
discuss the theoretical frame and then come back to the expectations at future
colliders.

\resection{BESS}

For a breaking of a group $G$ into a subgroup $H$
the Goldstones can be taken as the coordinates of $G/H$. We know that $H$
must contain $U(1)_{e.m.}$.  We need three Goldstones to give masses to
$W$ and $Z$. In addition we can guarantee for the standard parameter $\rho$ the
value
$\rho=1$ apart from weak corrections,
if we have a "custodial" $SU(2)$ \cite{19}.

The minimal $H$ in this case would have to
be $SU(2)$. In the SM the breaking is realized linearly with
scalars originally transforming as the $({1 \over 2}, {1 \over 2})$ of
$SU(2)_L\otimes SU(2)_R$. Such a direct product breaks into
 $SU(2)_{diagonal} $, with corresponding
breaking of $({1 \over 2}, {1 \over 2})$ into $1 \oplus 3$, describing the
physical Higgs and the 3 absorbed Goldstones.

The non-linear realization can
be seen classically as corresponding to the limit of infinite $m_H$. The
scalars can indeed be represented as proportional to a unitary matrix $U$.
In the formal limit $m_H \rightarrow \infty$ one is just freezing the
proportionality factor to the vacuum expectation value (called  $f$
to emphasize the formal similarity with $f_\pi$, the pion-decay constant
of strong chiral theory) and the scalar lagrangian is simply
$$
{\cal L} = {f^2 \over 4} Tr [(\partial_\mu U)(\partial ^\mu U^{\dagger)}]
$$
Such a lagrangian is obviously invariant under $SU(2)_L \otimes SU(2)_R$,
namely under $U\rightarrow g_L Ug_R^{\dagger}$ where $g_L, g_R$ belong to
$SU(2)_L, SU(2)_R$ respectively. The breaking into the diagonal $SU(2)$
is demanded by the (non-linear) unitarity condition $U^{\dagger}U=1$.

Before coming back to the our physical problem we want to summarize here some
general formal considerations. They have been known in different ways in
literature \cite{9}, \cite{10}, \cite{11}, \cite{12}.
We shall give here a systematic
presentation of the main relevant points.

Such considerations will be applied
in the following not only to the construction of the simplest original BESS,
but also to derive different extensions of BESS, that we shall also discuss and
examine, in view of possible phenomenological consequences.

Let us start by considering a local map $g(x)$ of the Minkowski space into a
compact connected Lie group $G$. One can from $g(x)$ construct the
Maurer-Cartan form $\omega$. The form $\omega$ is globally invariant under left
group multiplication.

Suppose $H$ is a connected subgroup of $G$, which we
shall identify as the ``unbroken subgroup". One has also for $H$ a local map
that we call $h(x)$.

We recall that $F_{\mu\nu}(\omega)$ vanishes by construction. One can decompose
$g(x)$ as a product
\be
g(x)= e^{iq(x)} h(x)
\label{1}
\ee
In eq. \ref{1} $q(x)$ has components along the generators $X_i$ of $G$ not
belonging to the Lie algebra of $H$: $q_i(x)=Tr(q(x)X_i)$. The $q_i(x)$ are the
coordinates of the non-linear realization.

Let us assume right-multiplication invariance under the local $H$. The
Maurer-Cartan form $\omega$ can be decomposed as
\be
\omega=\omega_{\parallel} +\omega_{\perp}
\label{2}
\ee
Parallel and orthogonal in \ref{2} is intended in relation to the unbroken
subgroup $H$.

Under the local right-multiplication one has
\be
\omega_{\parallel} \to h^\dagger \omega_{\parallel} h + h^\dagger \partial h
\ee
On the other hand
\be
\omega_{\perp} \to   h^\dagger \omega_{\perp} h
\ee

Under the restriction of the global $G$ into  the global $H$ the coordinates
$q_i(x)$ will in general transform non-linearly.

This construction can now be specialized to a symmetric space, in which case
one
adopts a standard basis for $Lie[G]$. In such a basis one has $[T_{\mu},
T_{\nu}]=if_{\mu\nu\lambda}T_{\lambda}$ for $T_{\mu} \in \; Lie[H]$, and
$[T_{\mu},X_i]=ig_{\mu i j}X_j$, $[X_i, X_j]=i g_{i j \mu}T_{\mu}$. Also we
recall that there exists in such a case a parity operation acting as an
automorphism of the algebra.

The standard non-linear realization of Callan, Coleman, Wess, and Zumino is
obtained by gauge-transforming $Tr(\omega_{\perp}^2)$ with $h^{-1} (x)$. One
obtains
\be
\omega_{\perp}=\left[ e^{-iq(x)}\partial e^{iq(x)} \right]_{\perp}
\ee
One can use the formula of Baker-Campbell-Hausdorf to write $\omega_{\perp}$ as
\be
\omega_{\perp}=i [adj(iq(x))]^{-1} sinh [adj(iq(x))]\partial q(x)
\ee
where the $adj$ of an operator is the adjoint defined in $Lie[G]$.

The standard results (current algebra, PCAC. etc.) are recovered by expanding
\be
f^2 Tr[\omega_{\perp}^2]= -f^2 Tr \left[ (\partial q(x))^2-\frac{1}{3} \partial
q(x) [q(x),[q(x),\partial q(x)]]+ \ldots \right]
\ee

The additional step is to introduce the gauge field $\eta$ of the local
unbroken subgroup $H$
\be
\eta \to h^\dagger \eta h +h^\dagger \partial \eta
\ee
One defines a covariant derivative (remark the arrow to the left)
\be
D g(x)=g(x) (\buildrel \leftarrow \over \partial +\eta)
\ee
For the field $\zeta$ defined as
\be
\zeta=\omega_{\parallel}-\eta
\ee
one has
\be
\zeta \to h^\dagger \zeta h
\label{4}
\ee
Eq. \ref{4} allows for a new symmetric term
\be
f'^2 Tr[\zeta^2]
\ee
with a new constant $f'$ which appears in addition to $f$ ($f$ is for instance
$f_{\pi}$ in QCD).

Now, $\eta$ remains as an auxiliary field as long as it does not develop a
kinetic term. This would give back the standard theory of non-linear
realizations. The physical assumption we shall make
is that such kinetic term arises in
the renormalized theory.

In fact nothing prevents its appearance within the
overall symmetry frame, and special examples may lead to suggest that it will
indeed appear \cite{12}. We shall in the following assume this to be the case.
Or, at least, construct our models as realizing such a possibility.

Let us now come back to the physical problem.

For a description of the scalar sector evidencing the hidden local symmetry of
the model one introduces local group elements $L(x)$, $R(x)$  belonging to
$SU(2)_L$ and $SU(2)_R$ respectively. Under global transformations of $SU(2)_L
\otimes SU(2)_R$ these group elements are multiplied to their left by the
corresponding global group transformation. In addition one can ask invariance
under right-multiplication by a local group element of the unbroken $SU(2)_V$.

The Maurer-Cartan form $\omega^{\mu} dx_{\mu}$, where
$\omega^{\mu}=(\omega^{\mu}_L,\omega^{\mu}_R)=(L^{\dagger}\partial^{\mu}L,
R^{\dagger}\partial^{\mu}R )$, is decomposed into the component
$\omega^{\mu}_{\parallel}$, parallel to the subgroup
$SU(2)_V$, and the component
$\omega^{\mu}_{\perp}$, orthogonal to $SU(2)_V$:
$\omega^{\mu}=(\omega^{\mu}_{\parallel})_a T^a_V +(\omega^{\mu}_{\perp})_a
T^a_A$, where $T_V^a$ and $T_A^a$ are the vector and axial vector generators of
$SU(2)_L \otimes SU(2)_R$.

One has
\be
(\omega^{\mu}_{\parallel})_a=Tr\left[\frac{1}{2}\tau^a
(L^{\dagger}\partial^{\mu}L +
R^{\dagger}\partial^{\mu}R)\right] , \; \;
(\omega^{\mu}_{\perp})_a=Tr\left[\frac{1}{2}\tau^a
 (L^{\dagger}\partial^{\mu}L - R^{\dagger}\partial^{\mu}R)\right].
\ee

The parallel Maurer-Cartan component transform under the local $SU(2)_V$
invariance according to such local invariance, whereas the orthogonal component
transforms as if such symmetry were only global (that is it develops no
inhomogeneous term under the gauge transformation). In addition one introduces
the $SU(2)_V$ gauge field: $\eta^{\mu}=(\eta^{\mu})_a T^a_V$.

One considers
terms which are invariant under the whole set of global $SU(2)_L \otimes
SU(2)_R$ transformations and local $SU(2)_V$ transformations, as defined. For
constructing an effective lagrangian one limits to terms containing at most two
derivatives and which are linearly independent. The simplest term is $-f^2 Tr
[ \omega^2_{\perp}]$. If one writes $U=L R^{\dagger}$ one can rewrite such
term as
\be
-f^2 Tr[ \omega^2_{\perp}]=-\frac{1}{4} f^2 Tr\left[
 (L^{\dagger}\partial^{\mu}L - R^{\dagger}\partial^{\mu}R)^2\right]=
\frac{1}{4} f^2 Tr \left[(\partial^{\mu}U)^{\dagger}(\partial_{\mu} U)\right]
\ee
showing that one has only reexpressed the $\sigma$-model in terms of
alternative degrees of freedom.

However the availability of the field
$\eta^{\mu}$ allows for the new term $-f'^2 Tr [(\omega_{\parallel}-\eta)^2]$,
with a
new independent vacuum value $f'$. On the other hand in absence of a kinetic
term for $\eta^{\mu}$ the field equations derived by adding the two terms would
simply lead back to the original non linear $\sigma$-model. The main physical
assumption is that $\eta^{\mu}$ becomes a dynamical field.

The gauging of $SU(2)_L \otimes U(1)$ only implies the substitution of the
ordinary derivatives with covariant left and right derivatives, acting on the
left or right group elements respectively:
\be
D_{\mu}^{(L)}L=\partial_{\mu}L+ W_{\mu}^{(0)} L-L \eta_{\mu} \; ,\;
D_{\mu}^{(R)}R=\partial_{\mu}R+ Y_{\mu} R -R \eta_{\mu}
\ee
where $Y_{\mu}=Y_{\mu} \tau_3 /2 $, $W_{\mu}^{(0)}=\vec{W}_{\mu}^{(0)}
\cdot \vec{\tau} /2 $, $\eta_{\mu}=\vec{\eta}_{\mu}
\cdot \vec{\tau} /2 $, and we have added a superscript zero to $W$ to allow
us later for use of the simpler symbols for the physical fields that will
emerge after mixing. Notice the different positioning of for instance
$W_{\mu}^{(0)}$ and $\eta_{\mu}$ in the covariant derivatives.

The final
lagrangian will contain the term built up from the transverse Maurer-Cartan
component, the term
corresponding to the field $(\omega_{\parallel}-\eta )$
 and the kinetic energies of the gauge bosons $W_{\mu}^{(0)}$, $Y_{\mu}$, and
also of $\eta_{\mu}$:
\bea
{\cal L} & = & -\frac{1}{4} f^2 Tr \left[
 (L^{\dagger}D^{(L)}_{\mu}L - R^{\dagger}D^{(R)}_{\mu}R)^2 \right]
-\frac{1}{4} f'^2 Tr\left[
 (L^{\dagger}D^{(L)}_{\mu}L + R^{\dagger}D^{(R)}_{\mu}R-\eta_{\mu})^2 \right]
\nn \\
& + & kinetic~ terms~ for~ the~ gauge~ fields
\eea

The Higgs mechanism gives masses to all gauge bosons, except for the photon.
All scalar degrees of freedom are absorbed. Formally one finds that one has to
perform the following gauge transformation $(\Omega=R L^{\dagger})$ :
\be
\vec{W}^{(0)}=\Omega^{\dagger} \vec{\tilde W} \Omega +\Omega^{\dagger}
\partial \Omega \; , \;
\vec{\eta}=R^{\dagger} \vec{\tilde V} R +R^{\dagger}
\partial R
\ee

Finally one performs a rescaling of the fields according to
$\tilde{W} \to g \tilde{W}$, $Y \to g' Y$, $2 \tilde{V} \to g'' \tilde{V}$
and after separate diagonalization of the $2 \times 2$ charged and $3 \times 3$
neutral sectors one derives the physical vector boson states  $W^{\pm}$,
$V^{\pm}$ and $A$, $Z^0$, $V^0$ with masses and mixing angles.

As we have said,
in absence of kinetic term for the gauge field of the hidden local symmetry,
one can only recover the original gauged non linear $\sigma$-model. The
rescaling we have performed for the field $\tilde{V}_{\mu}$ allows however  for
a different way of looking at such a limit. When $g'' \to \infty$ the limit is
again reobtained. Therefore $g'' \to \infty$ must lead back to the standard
electroweak theory. This indeed happens quite evidently from the expressions
for the masses and mixings \cite{bess}.

The fermions, quarks and leptons, are
those of the standard families, left-handed fermions $\psi_L$ and right-handed
fermions $\psi_R$. Under the local $SU(2)_V$ they are assumed to be singlets.
Their couplings are then uniquely determined
\be
{\bar \psi}_L i \gamma^{\mu} \left( \partial_{\mu} + \vec{W}_{\mu}^{(0)}
\frac{\vec{\tau}}{2} +\frac{1}{2} (B-L)Y_{\mu} \right) \psi_L +
{\bar \psi}_R i \gamma^{\mu} \left( \partial_{\mu} + \left(
\frac{\tau_3}{2} +\frac{1}{2} (B-L)Y_{\mu} \right)\right) \psi_R
\ee
from which one obtains the coupling constants.

An alternative procedure to discuss
 the "hidden gauge symmetries" for
a quite general model, that is perhaps more
related to usually employed notions, is only to enlarge the
initial symmetry and correspondingly enlarge the scalar sector and the
number of the needed non-linear conditions.

To obtain the  model of ref.
\cite{bess}
one adds to $SU(2)_L\otimes SU(2)_R$ a group $SU(2)_V$ and realizes
non-linearly the breaking $SU(2)_L\otimes SU(2)_R\otimes SU(2)_V\rightarrow
SU(2)_{diagonal} $. The Goldstones are six (coordinates of the quotient).
They are all absorbed, giving masses to $W, Z$  and $V$.

The Goldstones
are described by two unitary matrices $L$ and $R$, which transform as
$L \rightarrow g_L Lh$, $R \rightarrow g_R Rh$ where $g_L, g_R, h$ belong
to $SU(2)_L, SU(2)_R, SU(2)_V$ respectively. Forgetting the unitarity
conditions one would have $L, R$ transforming as $({1 \over 2},0,
{1\over 2})$ and $(0, {1\over 2}, {1\over 2})$ under
$SU(2)_L \otimes SU(2)_R \otimes SU(2)_V$. The unitarity conditions
$LL^\dagger = 1, RR^\dagger =1$ lead to the wanted breaking.

The procedure
then consists in writing down the most general Lagrangian with at most
two derivatives invariant under $SU(2)_L\otimes SU(2)_R\otimes SU(2)_V$ for
the unitary local matrices $L$ and $R$ and satisfying the symmetry
$L\leftrightarrow R$. One then introduces gauge fields for the subalgebra
$SU(2)_L\otimes U(1)_Y \otimes SU(2)_V$ and adds kinetic terms for them.
The related gauge couplings are $g,g'$, as usual, and $g''$  for $SU(2)_V$.

In the formal strong coupling limit, $g''\rightarrow \infty$, the kinetic term
of the $SU(2)_V$ gauge fields vanishes, and the fields become auxiliary.
Their elimination brings back to the non-linear formulation of the SM.

The model developed in ref. \cite{bess} contains the
massive dynamical gauge bosons corresponding to the "hidden" $SU(2)_V$
gauge symmetry. As we have said,
the importance of such vector bosons, in comparison to other
composite degrees of freedom, is that they can mix with $W$ and $Z$, and
therefore play an important role in phenomenology. A model  with one extra
triplet of vector bosons, based on $SU(2)_L \otimes SU(2)_V \otimes U(1)_Y$
and constrained by $\rho=1$ at tree level,
has also been considered \cite{schild}; BESS is obtained by specialization
of the parameter space of the model.

The same remarks however
would also apply to axial-vector bosons which also could mix. In the absence of
a complete dynamical treatment, which would enlighten us on the relative
role of vector and of axial bosons, one can only develop a general scheme
which contains both, and then discuss and compare the various phenomenological
predictions. From the point of view of our way of treating "hidden symmetries",
that is by adding the "hidden symmetries" at the start and then increasing
the number of scalar fields and of non-linear conditions, the inclusion of
the axial degrees of freedom appears indeed
as a natural and in principle very simple extension \cite{axial}.

One has to start from an initial symmetry $G$ consisting of a
global $ SU(2)_L \otimes SU(2)_R$ times a local $SU(2)_L \otimes SU(2)_R$
and introduce besides $L$ and $R$, transforming as $L \sim ({1 \over 2}, 0,
{1 \over 2}, 0)$ and $R \sim (0, {1\over 2},0, {1\over 2})$ under the above
sequence of groups, an additional $M$ transforming as $M\sim (0,0,{1\over 2},
{1\over 2})$.  The matrix $U=LM^\dagger R^\dagger$ will then transform only
globally and be unaltered by the local (hidden) group.
By imposing the non-linear, unitarity conditions $L^\dagger L =
R^\dagger R = M^\dagger M = {\bf 1}$ we realize again the breaking
from $G=[SU(2)_L\otimes SU(2)_R]_{global}\otimes [SU(2)_L\otimes
SU(2)_R]_{local}$ down
to the diagonal subgroup $H=SU(2)_{diagonal}$.

Altogether we have
9 Goldstone modes. At this point we can write down the most general
two-derivative lagrangian, invariant under the group $G\otimes P$, where $P$
parity transformation ($L \leftrightarrow R$ and $M \leftrightarrow
M^\dagger$), required to
make contact with the  non-linear $\sigma$-model limit of the standard model.

We first build up
covariant derivatives with respect to the local group:
\def \LL{{{\bf L}_\mu}}
\def \RR{{{\bf R}_\mu}}
\bea
& &D_\mu L =\partial_\mu L - L\LL\nn\\
& &D_\mu R =\partial_\mu R - R\RR\nn\\
& &D_\mu M =\partial_\mu M - M\LL + \RR M
\eea
where $\LL$ and $\RR$ are the Lie algebra valued gauge fields of
$({SU(2)_L})_{local}$ and $({SU(2)_R})_{local}$ respectively.

One now constructs the invariants of our original group extended by the
parity operation. One finds
\bea
{I}_1&=&\Tr ({L}^\dagger D_\mu {L}
       -M^\dagger D_\mu M-M^\dagger {R}^\dagger
          (D_\mu  {R}) M)^2\\
{I}_2&=&\Tr ({L}^\dagger D_\mu {L}+M^\dagger
           {R}^\dagger (D_\mu {R}) M)^2\\
{I}_3&=&\Tr ({L}^\dagger D_\mu {L}-M^\dagger {R}
         ^\dagger (D_\mu {R}) M)^2\\
{I}_4&=&\Tr (M^\dagger D_\mu M)^2
\eea

Using these invariants is it now possible to write down
 the most general Lagrangian with at
most two derivatives in the form:
\be
{\cal L}=-\frac{v^2}{16} (a {I}_1+b {I}_2+c {I}_3
+d {I}_4)+~kinetic~terms~for~the~gauge~
               fields
\ee
where $a,~b,~c,~d$ are free parameters and furthermore the gauge coupling
constant for the fields $\LL$ and $\RR$ is the same.

It is not difficult to see that this Lagrangian is the same
one would obtain from the hidden gauge symmetry approach
\cite {10}. The requirement of getting back
the non-linear $\sigma$-model in
the limit in which the gauge fields $\LL$ and $\RR$ are decoupled is
satisfied by imposing the following
relation among the parameters $a,b,c,d$
\be
a+ \frac{cd}{c+d}=1
\label{constr}
\ee

The gauging of the previous effective Lagrangian with respect to the standard
gauge group
$SU(2)_L\otimes U(1)_Y$ is obtained by the following substitutions:
\bea
D_\mu {L} &\to &  D_\mu {L} =
\dmu {L} -{L}
            (V_\mu-A_\mu)+ W_\mu {L}\\
D_\mu {R} &\to & D_\mu {R} = \dmu {R} -{R}
           (V_\mu+A_\mu)+ Y_\mu {R}\\
D_\mu M &\to & D_\mu {M} =\dmu M -M (V_\mu-A_\mu)+(V_\mu+A_\mu) M
\eea
where $V_\mu=(\RR+\LL)/2$ and $A_\mu=(\RR-\LL)/2$ are the fields describing the
new vector and axial-vector resonances.

Is it possible to fix the gauge so that $L=R=M=1$, obtaining the following
Lagrangian:
\bea
{\cal L} & = & -{v^2\over 4}
\left[a~tr(W-Y)^2+b~tr(W+Y-2 V)^2 +c~tr (W-B+2 A)^2 +\right. \nn \\
& + & \left. d~tr(2 A)^2 \right] + kinetic~terms~ for~ V_{\mu},
{}~A_{\mu},~W_{\mu},~Y_{\mu}
\label{lagr}
\eea

We still want to mention another, more intuitive,
approach to obtain the Lagrangian \ref{lagr}. One starts from the
transformation properties of the gauge fields $W^3_\mu$, $Y_\mu$,
$V^3_\mu$ and $A^3_\mu$ under the electromagnetic $U(1)_{em}$ gauge
transformations:
\bea
\delta W^3_\mu(x)&= &-{1\over g}\partial_\mu\lambda(x)\nn \\
\delta B_\mu(x)&= & -{1\over g'}\partial_\mu\lambda(x)\nn \\
\delta V^3_\mu(x)&= & -{2\over g''}\partial_\mu\lambda(x)\nn \\
\delta A^3_\mu(x)&=0
\label{gauge}
\eea
In the limit
$g'=0$, an $SU(2)$ global symmetry is defined,
under which $W$, $V$ and $A$
transform as triplets.

The lagrangian ${\cal L}$ is just that of
a massive Yang-Mills theory invariant under
the $U(1)_{em}$ gauge transformations given in eq. \ref{gauge} and
the "custodial" $SU(2)$.
The constraint given in eq. \ref{constr}
 is obtained by asking that in the limit
$g''\to\infty$ the lagrangian ${\cal L}$ reproduces the SM terms:
\be
{\cal L}_{SM}=-{v^2\over 4} tr(W-Y)^2+{\cal L}_{kin}(W,Y)
\ee

To summarize: the model describes the interactions of the
vector and axial-vector gauge bosons $A$ and $V$ of $[SU(2)_L\otimes
SU(2)_R]_{local}$
with the gauge bosons $W$, $Z$ and $\gamma$  of $SU(2)_L\otimes U(1)_Y$.
The original
9 Goldstone bosons are eaten up by $V$, $A$, $W$ and $Z$.

The parameters are:
$f$, $g$, $g'$; three independent coefficients in front of the lagrangian
invariants and the additional gauge coupling constants $g_A$ and $g_V$ of
the "hidden symmetry" group.

An important distinction from other schemes appears at this stage. Our
framework does not allow for vector resonant $SU(2)$ singlets, such as the
state which in the hadronic language would correspond to the $\omega$ meson, as
it instead naturally happens in schemes which try to mimic the QCD behavior.
The reason for this is that our starting $SU(2)_L \otimes SU(2)_R$ global
symmetry can be enlarged only by additional $SU(2)$ factors which, once gauged,
give rise to triplets of massive gauge bosons. On the other hand, if one tries
to gauge only a particular subgroup $U(1)$ of an extra $SU(2)$, unwanted
Goldstone bosons appear, with embarassing phenomenological consequences. It
would be different, of course, if the starting global symmetry were
$SU(3)_L \otimes SU(3)_R$. One thus expects, on merely symmetry grounds, that
technicolor, for instance, has general distinctive features with respect to the
scheme we have just described.

Couplings to fermions can be introduced following,
for instance, ref \cite{bess}. In a minimal choice, fermions couple to
the new vector bosons $V$ and $A$ only through the mixing of $V$, $A$ with
$W$ and $Z$.

The couplings among fermions and gauge bosons as well as
the low-energy ($\sqrt{s}\ll m_W$) charged and neutral currents lagrangian
can be straightforwardly derived.
One observes that $G_F=(\sqrt{2} f^2)^{-1}$, as in the SM, and, more
remarkably, $\rho =1$ at tree level. Therefore low-energy charged currents are
unaffected, whereas the neutral ones are only modified in the
expression for the Weinberg angle
and by the presence of an extra $j_{e.m.}^2$ term. Of course a sizeable
modification is given by the shift of the ordinary gauge boson masses.

The minimal chiral structure $SU(2)_L \otimes SU(2)_R$ of the original BESS can
be easily extended to a larger $SU(N)_L \otimes SU(N)_R$.

The most apparent
feature, in such a case, is the appearance of spin-zero pseudogoldstones,
due to the spontaneous breaking of the global $SU(N)_L \otimes SU(N)_R$
to the diagonal $SU(N)_V$. They are therefore $N^2-1$; three of them give mass
to the $W$ and $Z$. The others in general will not remain massless, due to the
interactions explicitly breaking the global symmetry group. For instance the
standard model gauge interactions contribute to the pseudo-Goldstone mass
spectrum \cite{peskin}.

It is however clear that other
interactions explicitly breaking the global symmetry group
 must also be present and taken into
account. We are here referring in particular to the mechanism which is
 responsible for the generation of the masses of the ordinary fermions.

 If for instance
we think to an extended technicolor scheme, the gauge interactions associated
to the generators connecting ordinary fermions to technifermions will in
general break the chiral symmetry $G$, which in this model is related to the
technifermion sector. Since the interactions considered are those responsible
for the generation of the fermion masses, it is natural to expect that the
induced pseudo-Goldstone masses are somehow related to the fermionic mass
spectrum. A quantitative analysis is presented in \cite{pseudo}.

Extended BESS contains explicit vector and axial-vector resonances. The
phenomenology of ordinary technicolor, in its low energy limit, would
correspond to a specialization of extended BESS.

The simplest construction for extended BESS
uses a local copy of the global chiral symmetry and goes through classification
of the relevant invariants, as shown before for $G=SU(2) \otimes SU(2)$.
The same results follow from the hidden gauge
symmetry approach. The standard electroweak $SU(2)\otimes U(1)$ and $SU(3)$ are
gauged and a definite  mixing scheme emerges for the gauge bosons and the
vector and axial-vector resonances. The physical photon and the physical gluon
remain automatically massless and coupled to their conserved currents.

The quantitative estimates have been restricted to the "historical" case $N=8$,
although a number of results are more general. Through their mixing with the
gauge bosons of $SU(2)_L \otimes U(1) \otimes SU(3)_c$, some of the vector and
axial vector resonances acquire a coupling to quarks and leptons, and are thus
expected to be produced at proton-proton and electron-positron colliders of
sufficient high energy. In $SU(8)$ these spin-1 bosons are an $SU(2)$ vector
triplet and axial triplet, an overall singlet, and a vector color octet, the
last one susceptible to be produced through the stronger color interaction.

The effective charged current-current interaction of extended BESS reproduces
the SM interaction, after identification of the relevant scale parameter with
the square root of the inverse Fermi coupling. Also,
for any chiral $SU(N)_L \otimes
SU(N)_R$, it can be seen that the neutral current-current interaction strength
corresponds to a $\rho$-parameter of 1, because of the diagonal $SU(N)$ which
is supposed to remain unbroken. All these results are of course corrected by
radiative effects.

If one tries to compare $SU(8)$-BESS with the original
$SU(2)$-BESS one sees that one main difference, concerning low energy effective
interaction, lies in the role of the additional singlet vector-resonance,
mentioned above. In addition the extension has new features, notably the
appearance of pseudogoldstones.

\resection{BESS Phenomenology}

Testing the symmetry breaking mechanism of the electroweak interactions will be
one of the main task of future colliders. We have already discussed the
strategies for the Higgs search, and now we adress the topic of possible
signatures of a strong symmetry braking sector, taking BESS as a simple
parametrization containing all the relevant elements.

In its minimal version, the BESS model has three independent parameters; the
mass $M_V$ of the new triplet of vector bosons, their gauge coupling $g''$,
assumed to be much larger than $g$ and $g'$, and a parameter $b$ giving the
direct coupling of the $V$'s resonances to fermions.
The parametr $b$, even if it is free in an effective lagrangian approach,
can be thought of as generated by
radiative corrections and therefore is expected
to be small \cite{CK}.
 The Standard Model is
obtained in the limit $g'' \to \infty$ and $b=0$.

Ordinary gauge bosons ($W$
and $Z$) mix to the new vector bosons $V$ with a mixing angle of the order
$g/g''$, at least in the approximation $M_V>>M_W$. Due to this mixing the $V$'s
are coupled to fermions even in absence of a direct coupling ($b=0$).

It is important to notice that the mixing angle does not disappear in the
limit $M_V \to \infty$, but it has an asymptotic value $g/g''$: therefore there
is no decoupling, and for this reason the observables far from the resonance
 turn out to be quite
insensitive to the value of $M_V$ (at least in the limits of validity of the
model).

The new resonances from the strong symmetry breaking influence masses and
couplings of ordinary gauge bosons and couple to fermions. Therefore one
expects small deviations with respect to the Standard Model predictions already
at
collider energies far below the production thereshold $M_V$.

These small
virtual effects can be seen  in $e^+e^-$ colliders, where high-precision
measurements are possible. In absence of such deviations one can put bounds on
the parameter space of the model.

Of course if the mass $M_V$ of the new
resonances is below the maximal c.m. energy of the collider there will be a
peak in the $e^+e^-$ annihilation cross section; tuning the collider at an
energy $\sqrt{s} \approx M_V$ would provide for a $V$'s factory allowing to
measure
properties and couplings of the new particles. But one may expect
 to see dominant
peaks below the maximum c.o.m. energy even without tuning the beam energies,
due to beamstrahlung.

If the new vector bosons are too heavy  to be produced as  resonances, one has
to look for deviations from the Standard Model values of the observables. At
LEP1, at the $Z$ resonance, the relevant couplings are those among the
$Z$ and the fermions, which enter in the process $e^+ e^- \to f {\bar f}$. In
BESS they differ from the SM ones up to terms proportional to $g/g''$ or $b$.
Furthermore, due to the mixing, also the values of the masses of $W$ and $Z$
bosons get shifted.

Putting together the data from LEP1 and CDF/UA2 on  the masses
of $W$ and $Z$, the widths of $Z$ into leptons and hadrons, and asymmetries,
one gets
severe restrictions on the parameter space of the model. This space is
essentially the plane ($b$,$g/g''$) because the observables are almost $M_V$
independent. For instance at $b=0$ one finds $g/g'' <0.06$.

Future $e^+ e^-$ linear colliders with different c.o.m. energies and
luminosities have been proposed; a collider with energy up to $500 \; GeV$
has concentrated most of the studies \cite{ee500}, but at the same time
possibilities of c.o.m. energies of 1 or 2 $TeV$ have been discussed.

In order
to test the hypothesis of a strongly interacting symmetry breaking sector the
channel $e^+ e^- \to W^+ W^-$ is particularly interesting, and large deviations
from the Standard Model predictions may be obtained. This is due to the strong
coupling between the longitudinal $W$ bosons and the new neutral resonance
$V^0$; furthermore in BESS the Standard Model cancellation among the
$\gamma$-$Z$ exchange diagrams and the neutrino contribution is destroyed.
Therefore the differential cross section grows  with the energy .

However, explicit calculations show that the leading term in $s$ is suppressed
by a factor $(g/g'')^4$ and,  at the energies considered here, it is the
constant term of the order $(g/g'')^2$ that matters.

Final $W$ polarization reconstruction can be done considering one $W$ decaying
leptonically and the other hadronically \cite{fujii},
 and it is relevant to constrain the model,
even if already at the level of unpolarized cross section one gets important
restrictions. Assuming an integrated luminosity of $20 \; fb^{-1}$,
$\sqrt{s}=500 \; GeV$ and $b=0$, it is possible to improve the LEP1 limit on
$g/g''$ over the whole $M_V$ range if polarization is measured, up to $M_V
\approx 1 \; TeV$ for unpolarized $W$.

$W^+ W^-$ pairs can be produced also through a mechanism of fusion of a pair of
ordinary gauge bosons, each being initially emitted from an electron or a
positron. This potentially interesting process allows, for a given c.m.
energy, to study a wide range of mass spectrum for the $V$ resonance, but it
becomes important for energies bigger than $2 \; TeV$.

Measurements of the various observables (cross sections and asymmetries)  of
the fermionic channel $e^+e^- \to f {\bar f}$ will not give a real improvement
with respect to the existing bounds from LEP1. The most sensitive observables
are
the left-right asymmetries, which need polarized $e^+ e^-$ beams; but also in
this case the bounds improve only for $M_V$ close to the value of the collider
energy.

In conclusion, concerning $e^+ e^-$ colliders,
we can say that they could
give the possibility to study the neutral
sector of symmetry breaking; $V^0 - Z$ mixing, $V^0 f {\bar f}$ and $V^0 W^+
W^-$ couplings. As we will see below there is complementarity with respect to
$pp$ colliders (LHC),  allowing to explore $V^{\pm}$ resonances through the
decay channel $W^{\pm}Z$. At proton colliders, as mentioned before, the channel
$V^0 \to W^+W^-$ is difficult to study due to background problems, and $V^0 \to
l^+ l^-$ has a very low rate.

Proton-proton colliders, as LHC, have a great potentiality for discovering new
strong
interacting gauge bosons, but of course
they are not as clean as $e^+ e^-$ colliders; the
high hadronic jets background makes the signals difficult to analyze, and if
new particles are found their properties could not be investigated in detail.

At proton colliders there are two possible mechanisms to produce $V$
resonances; $q {\bar q}$ annihilation and $WW (WZ, ZZ)$ fusion. In the first
mechanism a quark-antiquark pair annihilates into a $V$, which decays mostly
into a
pair of ordinary gauge bosons because the couplings $V^0 W^+_L W^-_L$ and
$V^{\pm} W^{\mp}_L Z_L$  are strong (of the order $g''$). We stress that this
process of annihilation always takes place in BESS, even if $b=0$, due to the
mixing between ordinary and new gauge bosons.
 We notice that in BESS there is
no coupling $V^0 Z Z$.

The second mechanism goes through fusion of a pair of ordinary gauge bosons,
both of them initially emitted from a quark or antiquark leg, to give a $V$
resonance decaying into a pair $W^{\pm}Z$ or $W^+ W^-$. The cross section is
obtained by a double convolution of the fusion cross section with the
luminosities of the initial $W/Z$'s inside the quarks and the structure
function of the quarks inside the protons.
 In the $q {\bar q}$ annihilation
only the convolution with the structure functions of the quarks is needed.
The amplitude of the elementary fusion process is strong in BESS: in fact the
scattering of two longitudinally polarized $W/Z$'s proceeds via the exchange of
a $V$ vector boson with large couplings (of the order $g''$ at each vertex).

As we pointed out before, the interesting channel at proton colliders is
$pp \to W^{\pm}Z + X$, because the $W^+W^-$ channel has a strong background
from $pp \to t {\bar t}+X$ and the $ZZ$ one is not resonant in BESS.
Monte Carlo simulations have been performed on the channel $WZ$ \cite{Pauss},
considering only the leptonic decays of $W$ and $Z$. To isolate the signal from
the background, coming from Standard Model $WZ$ and $t {\bar t}$ production,
one requires three isolated leptons, two of them reconstructing a $Z$ with high
transverse momentum. At LHC, with c.o.m. energy of 16 $TeV$, it turns out that
to reach a 2 $TeV$ mass for the $V$ an integrated luminosity greater than
$10^5 \; pb^{-1}$ is needed.

So far we have discussed the effects of the triplet of vector
resonances $V$. The real situation could be more complex: axial-vector
resonances might modify in a relevant way the predictions of the minimal model
with only vectors \cite{axial}. As a general feature, virtual effects and
deviations from the Standard Model coming from the vector and axial-vector
sector tend to cancel each other, and the final physical effects depend on the
relative weight of the two contributions. In some region of the parameter space
of the model there could be complete cancellations and no deviations from the
Standard Model would be observed, at least at energies below the new
resonances. The discovery of a strong electroweak sector only through virtual
effects and precision measurements could therefore be difficult and ambiguos.
The direct discovery of new resonances at the $TeV$ scale would be in such a
case  determinant.

In the extended BESS model a richer phenomenology appears. There are $N^2-1$
vector and $N^2-1$ axial-vector new resonances, associated to the local copy of
the global $SU(N)_L \otimes SU(N)_R$. These resonances mix with the ordinary
gauge bosons: in the case $N=8$ the neutral gauge sector involves the mixing of
the fields $W^3, Y,V^3,A^3,V_D$. $V_D$ is a chiral singlet, and its mixing
makes the colorless gauge sector of $SU(8)$-BESS different from the model based
on $SU(2)_L \otimes SU(2)_R$. The $W^{\pm}$, $V^{\pm}$ and $A^{\pm}$ sector is
like in $SU(2)$-BESS. Concerning the colored sector, the $SU(3)_c$ gluons mix
with a color octet of vector resonances $V_8^{\alpha}$

Another new feature is of course the presence of pseudo-Goldstone bosons.
We will indicate with $P^{\pm}$ ($P^0$) the lightest charged  (neutral) ones,
discussing in the following possible signatures at future accelerators
\cite{pseudofeno}.

Linear $e^+e^-$ colliders give the possibility to study the production of pairs
of charged pseudo-Goldstone bosons. They can be produced at the $V$ resonance
through the process $e^+ e^- \to V \to P^+ P^-$. The main decay mode of a
charged $P$ is $P^+ \to t {\bar b}$, if the pseudo-Goldstone is heavy enough.
We have therefore to analyze the final state $P^+P^- \to t {\bar b} {\bar t}
b$,
and compare it with the background. There are three background sources: $e^+e^-
\to W^+ W^-$, $e^+ e^- \to ZZ$, $e^+ e^- \to t {\bar t}$ and they have been
already studied in the process of charged Higgs boson production \cite{ch}.
Tagging one $b$ in the final state easily reduce the background $e^+e^- \to
W^+W^-$, while the others two sources are smaller than the signal, at least in
a reasonable range of the model parameter space.

At LHC pseudo-Goldstone bosons can be produced from a decay of a $V$ resonance,
previously produced from quark-antiquark annihilation or from a fusion process.
The charged channel, $pp \to V^{\pm} \to P^{\pm} P^0 +X$, gives the signal
$t {\bar b} b {\bar b}$ or $t {\bar b} g g$, because $P^0$, the lightest
pseudo-Goldstone, decays mainly in $b {\bar b}$ and $gg$. For the neutral
channel $pp \to V^{0} \to P^{+} P^- +X$ one looks for the signal $t {\bar b}
{\bar t} b$. The backgrounds are expected to be large, and a careful study is
needed. In a study done for the case of charged Higgs boson pair at LHC
\cite{ch2} it has been shown that a good $b$ tagging is necessary to identify
the signal.

\resection{Conclusion}

The problem of electroweak symmetry breaking has acted in these last years as a
dominant stimulus for imagining new physics beyond the standard model.

In this contribution we have first reviewed the theoretical situation and the
different  perspectives on the problem. We have then shortly summarized
prospects at existing and future colliders, relevant to the question of
electroweak symmetry breaking.

In the main part of this work we have concentrated on BESS (Breaking
Electroweak Symmetry Strongly) as a simple scheme to describe an alternative
breaking scheme avoiding elementary scalars. We have discussed the mathematical
frame, according to two possible general constructions, the possible directions
for extensions, specialization to the technicolor phenomenology, and general
characteristic features.

Finally we have tried to summarize the work done to put limits on the BESS
parameters from presently available precision data, and the exploratory work
on BESS predictions for future colliders such as LHC and $e^+e^-$ linear
colliders at very high energy.

\vspace{1cm}
\noindent
{\bf ACKNOWLEDGMENT}\\

Many physicists (R. Casalbuoni, P. Chiappetta, M.C.
Cousinou, A. Deandrea, S. De Curtis, D. Denegri, D. Dominici, F. Feruglio,
A. Fiandrino, I. Iosa, B. Mele, G. Nardulli, F. Pauss, T. Rodrigo, P. Taxil, J.
Terron, and others) have contributed to different aspects of BESS studies. We
want to thank them for their valuable work.
\vspace{1cm}

%\newpage

\end{document}